\documentclass[12pts] {article}
\usepackage[utf8]{inputenc}
\usepackage{authblk}
\usepackage{setspace}
\usepackage[margin=1.25in]{geometry}
\usepackage{graphicx}
\graphicspath{ {./figures/} }
\usepackage{subcaption}
\usepackage{amsmath}
\usepackage{chemformula}
\usepackage{booktabs}
\usepackage{makecell}
\usepackage{tabularx}
\usepackage{longtable}
\usepackage{array}
\usepackage{siunitx}
\usepackage{comment}
\usepackage[version=4]{mhchem}
\usepackage{subcaption}
\usepackage{tcolorbox}
\usepackage{siunitx} 
\usepackage{soul}
\usepackage{float}

\usepackage[
  maxbibnames=99,
  backend=biber,
  style=numeric,
  sorting=none,
  natbib=true,
  doi=true,
  url=false,
  isbn=false
]{biblatex}
\title{Structural, Energetic, Electronic, and Vibrational Properties of Boron-Substituted Tungsten Clusters}

\author[1]{Akshata M. Waghmare}
\author[2]{Gautam S. Kamat}
\author[1]{Sajeev S. Chacko}
\author[1]{Chetan V. Gurada}
\author[1]{Balasaheb J. Nagare\thanks{Corresponding author}}

\affil[1]{Department of Physics, University of Mumbai,Kalina Campus, Santacruz (East), Mumbai 400098, Maharashtra, India}
\affil[2]{Department of Physics, Chikitsak Samuha's Sir Satam and Lady Shantabai Patkar College of Arts and Science and V.P. Varde College of Commerce and Economics, Goregaon (West), Mumbai 400061, Maharashtra, India.}

\begin{document}
\def \wxby {W$_n$B$_m$}
\def \wthree {W$_3$}
\def \wtwo {W$_2$}
\def \wfour {W$_4$}
\def \wfive {W$_5$}
\def \wsix {W$_6$}
\def \wthirteen {W$_13$}
\def \wtwobone {W$_2$B$_1$}
\def \wthreebone {W$_3$B$_1$}
\def \wfourbone {W$_4$B$_1$}
\def \wfivebone {W$_5$B$_1$}
\def \wsixbone {W$_6$B$_1$}
\def \wthreebone {W$_3$B$_1$}
\def \wonebtwo {W$_1$B$_2$}
\def \wtwobtwo {W$_2$B$_2$}
\def \wthreebtwo {W$_3$B$_2$}
\def \wfourbtwo {W$_4$B$_2$}
\def \wfivebtwo {W$_5$B$_2$}
\def \wsixbsix {W$_6$B$_6$}
\def \wonebthree {W$_1$B$_3$}
\def \wtwobthree {W$_2$B$_3$}
\def \wthreebthree {W$_3$B$_3$}
\def \wonebfour {W$_1$B$_4$}
\def \wtwobfour {W$_2$B$_4$}
\def \wthreebfour {W$_3$B$_4$}
\def \wfourbfour {W$_4$B$_4$}
\def \wonebfive {W$_1$B$_5$}
\def \wtwobfive {W$_2$B$_5$}
\def \wthreebfive {W$_3$B$_5$}
\def \wfourbfive {W$_4$B$_5$}
\def \wfivebfive {W$_5$B$_5$}
\def \wonebone {W$_1$B$_1$}
\maketitle

\begin{abstract}
We report a study of the density functional theory of the structural, electronic and vibrational properties of small tungsten-boron clusters using the B3LYP exchange-correlation functional together with the QZVP basis set. A large number of possible isomeric structures were generated using the USPEX code interfaced with Gaussian~03, and the lowest-energy configurations were selected for detailed analysis. The results show that the addition of boron significantly modifies the geometry of tungsten clusters by reducing their symmetry and leading to the formation of shorter and stronger W-B and B-B bonds. With increasing boron concentration, boron atoms preferentially occupy edge and outer positions and gradually form interconnected B-B links within the clusters. These structural modifications improve the overall stability of the clusters, as reflected by the increase in binding energy, HOMO-LUMO energy gap, ionization potential, and chemical hardness with increasing boron concentration. The eigenvalue spectra show a greater separation between the occupied and unoccupied electronic states after boron substitution, indicating enhanced electronic stabilization and stronger electron localization. Vibrational analysis further reveals a gradual shift from low-frequency W-W vibrational modes in pure tungsten clusters to higher-frequency W-B and B-B stretching modes in boron-rich clusters, indicating an increase in the covalent character of bonding. Overall, the results show that boron improves the stability of tungsten clusters by modifying their bonding and electronic properties at the atomic scale.
\end{abstract}

\section{Introduction}
Transition–metal borides (TMBs) are refractory materials characterized by exceptional hardness, high melting temperatures, and chemical stability.
These remarkable properties originate from the coexistence of strong covalent B–B networks together with metallic bonding between boron and transition-metal atoms.~\cite{gu2022synthesis,moscicki2020influence,fuger2022anisotropic,fuger2019influence}.
Consequently, compounds such as TiB$_2$, MoB$_2$, and WB$_2$ are widely used in wear-resistant and high-temperature applications\cite{chertovskikh2021study,windsor2021tungsten}.
Recent advances in synthesis techniques, including thermal plasma processing and mechanochemical methods, have further improved their performance by the formation of nanostructured phases with enhanced stability and tunable physical properties.~\cite{radziejewska2020characterization,moscicki2021properties,kim2019synthesis}.
Moreover, alloying and defect engineering in systems such as ((W,Ti)B$_2$, WB$_{2-z}$, and Ta-alloyed WB$_2$ significantly enhance hardness, oxidation resistance, and thermal stability. 
These modifications also expand the functionality of transition-metal borides toward catalytic and energy-related applications, including Mo$_{1-x}$W$_x$B$_2$ electrocatalysts and WB-based heterostructures for lithium–sulfur batteries.~\cite{psiuk2023mechanical,park2019high,wu2026facilitated}.
Overall, the properties of TMBs can be systematically tuned through compositional design and nanoscale engineering.

Among the various TMB systems, tungsten borides have attracted particular interest due to their exceptional hardness, structural complexity, and potential applications in advanced mechanical and energy-related technologies~\cite{levine2009advancements}.
Due to their unique features, they have been used in numerous applications, such as cutting tools, wear-resistant coatings\cite{liu2025understanding,mishigdorzhiyn2020microstructure}  , abrasives\cite{artamonov1966abrasive}, corrosion-resistant materials, and electrode materials\cite{zhang2023tungsten,hatipoglu2022design}.
They are especially used for high-temperature and shielding applications\cite{erdogan2025enhanced,windsor2021tungsten,liu2023boriding,windsor2022activation,lin2022oxidation}.
These materials offer several advantages, such as low-cost tungsten and boron, reduced production costs due to higher boron content, and lower density, making them ideal for weight-sensitive applications.
However, the widespread use of transition metal borides is challenging because of their complex crystal structure\cite{cumberland2005osmium,qin2008rhenium}.

Several stoichiometric compositions of tungsten borides have been investigated so far: W$_2$B, WB, WB$_2$, W$_2$B$_5$ and WB$_4$~\cite{kiessling1947crystal,armas1973chemical,itoh1987formation,qin2018effect}.
Duschanek {\em et al}. studied the W$_{1-n}$B$_3$ phase by optimizing a set of parameter values describing the Gibbs energy of individual phases using thermodynamic models.
Additionally, they calculated the phase diagram, thermodynamic parameters of this system and compared it with experimental data~\cite{duschanek1995critical}.
Numerous recent theoretical studies have focused on exploring the stability of potential novel phases and their corresponding physical properties~\cite{zhong2013phase,zhao2010phase,chen2008electronic,zhang2017first}.
Zhao {\em et al}. conducted a comprehensive exploration of the W-B system across a pressure range of 0 to 40 GPa.
This not only confirmed all known stable compounds but also led to the discovery of two novel stable phases (P42$_1$m-WB and P2$_1$/m-W$_2$B$_3$) and three promising near-stable candidates (R3m-W$_2$B$_5$, Ama2-W$_6$B$_5$, and Pmmn-WB$_5$) at ambient pressure and $0$~K.
This study provides significant information on the complex crystallography of tungsten borides, identifying WB$_2$, WB$_4$ and WB$_5$ as promising candidates for advanced mechanical applications~\cite{zhao2018unexpected}.
The findings by Kvashnin {\em et al}. of a potentially superhard material, uncovered an unexpected link between the theoretically projected WB$_5$ and the experimentally observed WB$_4$.
The new material has a crystal structure similar to WB$_5$, with varying chemical compositions denoted by the formula WB$_{5-n}$ due to disorder and non-stoichiometry.
Their calculations identified average crystal structures matching experimental data, highlighted preferred local atomic arrangements, and elucidated specific atomic patterns.
These models enable detailed property calculations, facilitating comparison with experimental results~\cite{kvashnin2020wb5}.
In addition to bulk phases, finite tungsten and tungsten–boride clusters have also been investigated experimentally and theoretically.
Small W$_n$B$_m$ clusters can be considered as local "building blocks" of the bonding motifs that also appear in bulk tungsten borides like WB, WB$_2$, WB$_4$, and WB$_5$, the same W-W, W-B, and B-B units that stabilize small clusters form the extended networks that give bulk WB phases their hardness, high melting point, and chemical stability.
Hence, systematic studies on finite tungsten clusters~(W$_n$) would establish a clear understanding of how their structures and electronic properties would affect the properties of their bulk counterparts.
Early density functional theory~(DFT) investigations on very small clusters~($n$=2–4) focused on identifying stable geometries, spin states, and suitable relativistic treatments for tungsten~\cite{wang2005geometries,zhang2005density}.
Subsequent size-dependent analyses extended to larger systems, showing that W$_n$ clusters (upto $n$=16) gradually transform from planar or quasi-planar motifs into three-dimensional polyhedral structures, accompanied by systematic trends in binding energies, magnetic moments, and HOMO-LUMO gaps~\cite{du2009geometrical}.
A broader first-principles survey covering $n$=3–27 further explored global and local minima and introduced a jellium-like electronic shell model to rationalize the appearance of electronic "magic numbers" and enhanced stability at specific sizes~\cite{lin2008first}.
Studies on neutral and charged clusters ($n$=3–6) also simulated photoelectron spectra and vertical detachment energies, strengthening the link between theory and experiment~\cite{yong2009first}.
Later methodological advances, including basin-hopping and global optimization techniques, enabled more reliable exploration of the complex potential energy surfaces of tungsten nanoclusters~\cite{lin2014structural}.
Collectively, these works provide a solid reference framework for understanding structural evolution, electronic stabilization, and the onset of metallic character in pure tungsten clusters.

Building on this foundation, finite tungsten–boron (W$_m$B$_n$) clusters began to receive attention, although the literature remains comparatively limited.
The first explicit DFT studies on small W$_m$B$_n$ clusters with $m+n \leq 7$ examined their structures and relative stabilities~\cite{Xiuron2013DensityFT}, followed by investigations of polarizabilities and dipole moments to probe their response properties and charge distribution trends~\cite{yin2013theory}.
These works represent the earliest systematic attempts to treat bare WB clusters as distinct finite systems rather than extensions of bulk borides.
Although finite W and W-B clusters have been reported, a systematic investigation of boron substitution effects within homologous tungsten cluster frameworks remains largely unexplored.
These clusters could form novel superhard materials used as advanced catalysts and radiation-resistant components beyond bulk forms.

In the present work, we performed a comprehensive theoretical investigation of \wxby~clusters using density functional theory to gain insight into how the addition of boron atoms influences the structural stability, electronic characteristics, and bonding nature of tungsten at the nanoscale.
This approach aims to bridge the understanding between isolated atomic aggregates and bulk behavior in \wxby~systems.

This paper is organized as follows.
In section~\ref{sec:comp}, we describe the computational details and methodology used in this work.
In section~\ref{sec:results}, we highlight the findings of this work and provide detailed explanations.
We conclude this work in section~\ref{sec:concl} which summarizes the main findings and provides final insights.

\section{Computational Methodology}
\label{sec:comp}
To begin with, the structural search for all $W_nB_m$ clusters ($n,m \leq 5$) was carried out using the Universal Structure Prediction Evolutionary Xtallography (USPEX) code interfaced with the Gaussian~03 software package~\cite{g03}. For each cluster composition, nearly 2000 candidate geometries were generated to ensure broad sampling of the configurational space and to identify energetically favorable isomeric structures. The generated structures were subsequently optimized using density functional theory (DFT) calculations employing the B3LYP exchange-correlation functional together with the QZVP basis set. The B3LYP functional, which combines Becke’s hybrid exchange functional~\cite{becke1993density} with the Lee-Yang-Parr correlation functional~\cite{lee1988development}, provides a reliable description of electron correlation effects, while the QZVP basis set offers an accurate treatment of both tungsten and boron atoms~\cite{weigend2005balanced}. The resulting isomers were arranged in increasing order of their total energies, and nearly identical structures were removed using an energy tolerance of 10$^{-2}$~eV. The lowest energy structures are shown in Figure~\ref{fig:gs}.

The lowest-energy structures obtained from this procedure were selected for further analysis. To ensure reliable and unbiased geometries, all optimizations were performed without imposing any symmetry constraints, and the optimization process was continued until the maximum force on each atom became less than 1.5$\times$10$^{-5}$~Hartree/Bohr. The stability of the optimized clusters was further verified through harmonic vibrational frequency calculations performed at the same level of theory. The absence of imaginary frequencies confirmed that the obtained geometries correspond to true minima on the potential energy surface.

These validated low-energy structures were then used to study the structural, electronic, and vibrational properties of the $W_nB_m$~clusters. In particular, Raman spectra and eigenvalue spectra were calculated to understand their vibrational and electronic behavior. In addition, ionization potential, chemical hardness, and reactivity indices such as Fukui functions were analyzed to gain further insight into the stability and chemical reactivity of the clusters.

\section{Results and Discussion}
\label{sec:results}

In this section, we discuss how the addition of boron atoms changes the structural, electronic, and vibrational properties of tungsten clusters. The results show that boron strongly influences the bonding pattern, geometry, and stability of the clusters. By comparing pure tungsten clusters with boron-substituted systems, we analyze how the bonding gradually changes from mainly metallic interactions in pure tungsten clusters to stronger and more directional bonding in boron-rich clusters. We also discuss how these structural changes affect the electronic stability, reactivity, and vibrational behavior of the clusters.

\subsection{Structural Properties}
For the sake of completeness, we first discuss the structural characteristics of the corresponding dimers, which provide useful insight into the bonding trends in larger clusters.
The B-B, W-B, and W-W dimers exhibit equilibrium bond lengths of 1.61, 1.96, and 1.99~\AA, respectively, with corresponding vibrational frequencies of 1004, 731, and 404~cm\(^{-1}\)~\cite{nist_b2,merriles2021chemical,hu1992optical,borin2010electronic}. 
The slightly shorter bond length may be related to the known tendency of hybrid functionals such as B3LYP to overbind transition-metal dimers\cite{yanagisawa2000investigation,barden2000homonuclear}.
The trend in bond strength follows B-B $>$ W-B $>$ W-W.
\begin{figure}[H]
    \centering
    \includegraphics[width=0.6\textwidth]{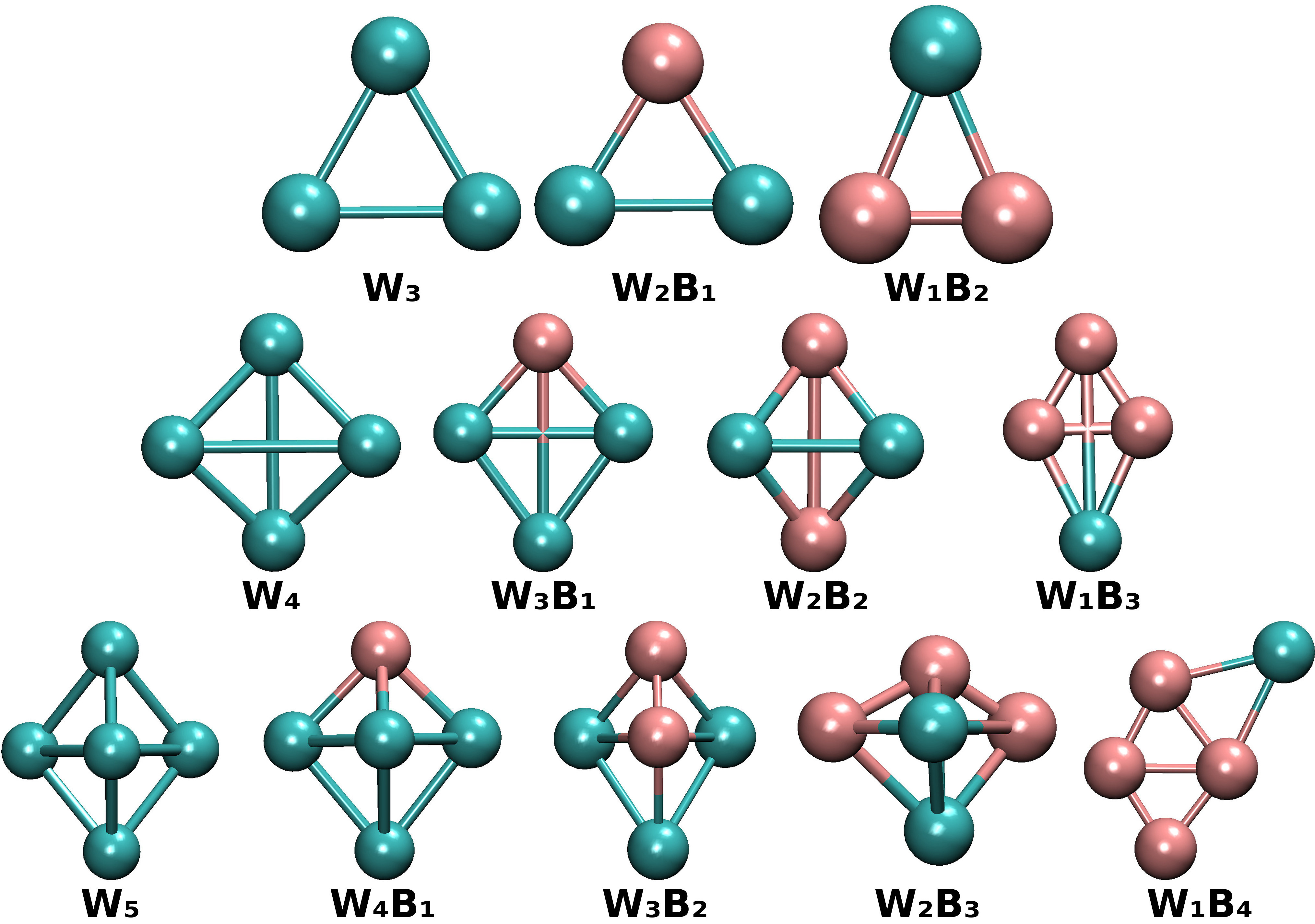}
    \caption{Optimized geometries of W$_n$B$_m$ clusters. Cyan and pink spheres represent tungsten (W) and boron (B) atoms, respectively.}
    \label{fig:gs}
\end{figure}
The optimized geometries of the \wxby\ clusters are presented in Fig.~\ref{fig:gs}.
The W$_3$ cluster forms a triangle, W$_4$ forms a buckled square, and W$_5$ develops into a three-dimensional~(3D) polyhedral configuration.
A tungsten atom's natural tendency to maximize coordination through metal interactions is thus reflected in this shift from planar to 3D~structures.
With the addition of boron, significant modifications in the geometries are observed.
The smallest mixed system (WB) adopts a linear geometry.
As the number of boron atoms increases, (W$_2$B$_1$, W$_1$B$_2$), the boron atoms tend to occupy edge or terminal positions.
This preference is associated with the lower coordination of boron atoms and the higher electronegativity of boron.
For intermediate clusters such as W$_3$B$_1$, W$_2$B$_2$, and W$_1$B$_3$, the distortions in their structures are higher compared to their pure tungsten counterparts.
The substitution of tungsten by boron reduces the overall symmetry and leads to shorter bond lengths, particularly for W-B and B-B bonds.
In larger clusters (W$_4$B$_1$, W$_3$B$_2$, W$_2$B$_3$, and W$_1$B$_4$), boron atoms increasingly form B-B linkages and substructures, while the tungsten atoms act as a supporting metallic framework.
The structures become increasingly distorted and interconnected with increasing boron content.
Overall, boron substitution leads to (i)~reduced symmetry, (ii)~shorter interatomic distances, and (iii)~significant modification of the cluster geometries. 
These structural changes play a key role in enhancing the stability and modifying the electronic behavior of tungsten-boron clusters.
\subsection{Electronic properties}
\label{subsec:elect}
Next, we focus on the electronic properties and stability of the W$_n$B$_m$ clusters. The effect of boron addition is analyzed using binding energy, electronic energy levels, and reactivity parameters. These results help us understand how boron changes the bonding, stability, and overall behavior of tungsten clusters.

The stability of each \wxby\ cluster can be understood by analyzing the binding energy per atom, defined as:
\begin{equation}
E_{\text{bind/atom}} =
-\frac{E_{\text{tot}}(\text{\wxby})
- \left(nE_\text{W}^{\text{iso}} + mE_\text{B}^{\text{iso}}\right)}
{n + m}.
\end{equation}
Here, $n$ and $m$ represent the numbers of tungsten and boron atoms, respectively.
$E_\text{W}^{\text{iso}}$ and $E_\text{B}^{\text{iso}}$ are the total energies of isolated tungsten and boron atoms, and $E_{\text{tot}}(\text{\wxby})$ is the total energy of the optimized cluster. 
The binding energy gives a direct measure of the stability of a cluster.
A higher value of the binding energy indicates stronger bonding and greater thermodynamic stability relative to the separated atoms.

Table~\ref{tab:BE} shows the binding energy per atom (B.E./atom) values for all the studied clusters.
From these results, it may be observed that the binding energy increases upon introduction of boron atoms into the tungsten clusters. 
This clearly shows that boron stabilizes the system.
The W–B bonds are stronger than the W–W bonds, leading to a rearrangement of atoms to maximize these bonds.
Among the studied clusters, \wthreebtwo~exhibits the highest binding energy per atom, indicating it to be the most stable cluster with enhanced thermodynamic stability.

\begin{table}[H]
\caption{Optimized bond lengths (in~\AA) and average binding energy per atom (B.E./atom, in~eV) for \wxby~clusters.}
\resizebox{0.95\linewidth}{!}{%
\begin{tabular}{llllc}
\toprule
\textbf{Cluster} & \multicolumn{3}{c}{\textbf{Bond-length~(\AA)}} & \unit{\textbf{B.E/atom}~\textbf{(eV/atom)}} \\
\cmidrule{2-4}
 & \textbf{W-W} & \textbf{W-B} & \textbf{B-B} &  \\
\midrule
\textbf{\wthree} & 2.28 & - & - &  2.71\\
\textbf{\wtwobone} & 2.18 & 2.07,2.05 & - & 2.87\\
\textbf{\wonebtwo} & - & 1.96 & 1.53 &2.90 \\
\hline

\textbf{\wfour} & 2.28 & - & - & 3.14 \\
\textbf{\wthreebone} & 2.25,2.65 & 2.00, 3.31 & - &  3.41   \\
\textbf{\wtwobtwo} & 2.29 & 2.01 & 2.94 &  3.38\\
\textbf{\wonebthree} & - & 1.98 & 1.60,1.71 &3.36 \\
\hline

\textbf{\wfive} & 2.37 & - & - &  3.57\\
\textbf{\wfourbone} & 2.35,2.35,2.61,2.37,2.95 & 2.09,2.22,2.09,2.31 & -  & 3.66 \\
\textbf{\wthreebtwo} & 2.27,2.40 & 2.14,2.30,2.02 & 1.62 & 3.80    \\
\textbf{\wtwobthree} & 2.31 &2.00,2.28,2.15,1.96,2.06 & 1.58,1.75 & 3.66   \\
\textbf{\wonebfour} & - & 2.02,2.07 & 1.52,1.57,1.61,1.75,1.79 & 3.67 \\
\bottomrule
\end{tabular}
}
\label{tab:BE}
\end{table}

To gain deeper insight into the electronic structure and stability of the clusters, we further examine their eigenvalue spectra.
\begin{figure}[H]
    \centering
    \includegraphics[width=\columnwidth]{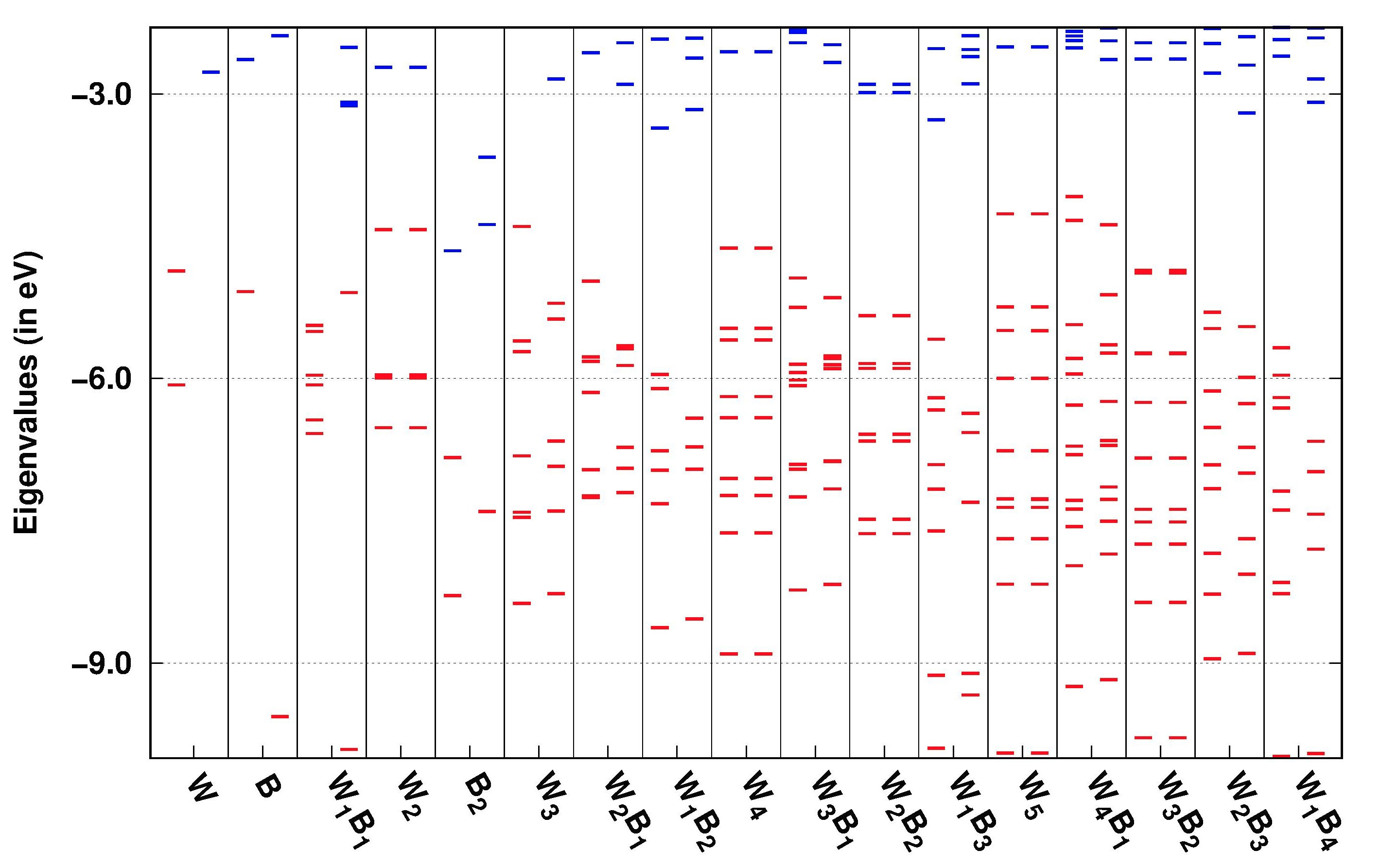}
    \caption{Eigenvalue spectra of \wthree, \wfour, and \wfive~clusters with boron substitution.
    Red and blue show occupied and unoccupied orbitals, respectively.}
    \label{fig:ev2}
\end{figure}

To gain further insight into the electronic structure of \wxby~clusters, we analyzed their eigenvalue spectra (see Fig.\ref{fig:ev2}).
The red and blue levels correspond to occupied and unoccupied orbitals, respectively, with two columns representing the $\alpha$- and $\beta$-spin components.
\wonebone, \wtwo, W$_3$, \wtwobone,  W$_3$B$_1$, \wtwobtwo,  and W$_4$B$_1$~exhibits some nearly degenerate orbitals.
whereas some pure tungsten~(W$_2$, W$_4$, W$_5$) and doubly sustituted boron~(W$_2$B$_2$ and W$_3$B$_2$) clusters are singlet in nature. Hence, these clusters exhibit non-magnetic character.
Furthermore, boron substitution significantly modifies this picture. As listed in Table~\ref{tab:cdft}, the boron-rich clusters W$_1$B$_2$, W$_1$B$_3$ and W$_1$B$_4$ show the largest orbital separation. This indicates comparatively higher electronic stability among the studied clusters.
A similar behavior has been reported in cluster systems where structural distortion and mixing of orbitals lead to splitting of energy levels and improved stability~\cite{sharma2022dft,}.

The electronic stability of the clusters can also be understood from their ability to lose or gain electrons. Therefore, we next analyze the vertical ionization potential (VIP) and vertical electron affinity (VEA) of the clusters.
The energies of the cationic and anionic \wxby~clusters were calculated without optimizing their structures to compute the VIP and VEA, respectively, that can serve as key descriptors of their stability and reactivity.
They were determined as:

\begin{equation}
\text{VIP} = E_{\mathrm{cation}} - E_{\mathrm{neutral}}
\end{equation}
\begin{equation}
\text{VEA} = E_{\mathrm{neutral}} - E_{\mathrm{anion}}
\end{equation}
where $E_{\mathrm{neutral}}$, $E_{\mathrm{cation}}$, and $E_{\mathrm{anion}}$ represent the total energies of the neutral, cationic, and anionic clusters, respectively, evaluated at the optimized geometry of the neutral cluster.
The calculated values are summarized in Table~\ref{tab:VIP-VEA}.

\begin{table}[H]
\footnotesize
\centering
\caption{Calculated vertical ionization potentials (VIP) and vertical electron affinities (VEA) of the \wxby~clusters. All values are reported in eV and correspond to the optimized ground-state geometries obtained from DFT calculations.}
\begin{tabular}{lrrrrcc}
\toprule
\textbf{Cluster} &
\multicolumn{3}{c}{\textbf{VIP}} &
\multicolumn{3}{c}{\textbf{VEA}} \\
\cmidrule(lr){2-4} \cmidrule(lr){5-7}
 & \textbf{This work} & \textbf{Expt} & \textbf{\% error}
 & \textbf{This work} & \textbf{Expt} & \textbf{\% error}
 \\
\midrule
\textbf{W$_2$} &6.69 &6.57 &1.8 &1.07 &1.46&26.71\\
\textbf{\wonebone} &7.86 & -&- &1.09 & -&- \\
\hline\\
\textbf{\wthree} &6.98 &6.05 & 15.37&1.35 &1.44&6.25 \\
\textbf{\wtwobone} &7.51 &- &- & 0.99&- &-\\
\textbf{\wonebtwo} & 8.54 &- &- &0.71 & -&-  \\
\hline\\
\textbf{\wfour} & 6.06 &5.61 &8.00 &1.19 & 1.64& 27.43 \\
\textbf{\wthreebone} & 6.59 & -&- & 1.23 &- &-\\
\textbf{\wtwobtwo} &7.18 & -& -&1.31 & -&- \\
\textbf{\wonebthree} & 8.41 &- & -&1.01 & -&- \\

\hline\\
\textbf{\wfive} &5.61 & 4.99& 12.42&1.32&1.58 &16.45 \\
\textbf{\wfourbone} &5.77  &- &- &1.28 &- &-\\
\textbf{\wthreebtwo} &6.48 & -&- & 1.18 &- &-  \\
\textbf{\wtwobthree} &7.17 & -& -&1.61 &- &-\\
\textbf{\wonebfour} &8.58 &- & -&1.56 &- &-\\
\bottomrule
\end{tabular}
\label{tab:VIP-VEA}
\end{table}

For pure tungsten clusters (W$_n$), the VIP values decrease slightly with increasing cluster size, from 6.98~eV for W$_3$ to 5.61~eV for W$_5$. 
The corresponding VEA values increases moderately from 1.19 to 1.35~eV, indicating a limited ability to accommodate additional charge.
With boron substitution, a systematic increase in the VIP values is observed across all cluster sizes.
For example, in the three-atom series, the VIP increases from 6.98~eV (W$_3$) to 7.51~eV (W$_2$B$_1$) and further to 8.54~eV (W$_1$B$_2$). 
A similar trend is observed for larger clusters, where VIP increases from 6.06~eV (W$_4$) to 8.41~eV (W$_1$B$_3$), and from 5.61~eV (W$_5$) to 8.58~eV (W$_1$B$_4$). 
Whereas, the VEA decreases with increasing boron substitution (from 1.35~eV (W$_3$) to 0.99~eV (W$_2$B$_1$) and 0.71~eV (W$_1$B$_2$)), reflecting reduced electron acceptance in tungsten-rich configurations. 
However, with increasing boron concentration, the VEA increases again, reaching a value of 1.31~eV for W$_2$B$_2$, and further increases to 1.56~eV for W$_1$B$_4$.
In particular, W$_2$B$_3$ has a relatively high VEA of 1.61~eV, which indicates enhanced electron affinity in boron-rich clusters.
Overall, table~\ref{tab:VIP-VEA} demonstrates that VIP increases monotonically with boron content, while VEA exhibits a composition-dependent trend, initially decreasing and then increasing in boron-rich clusters.
The increase in VIP reflects stronger electron binding, whereas the eventual rise in VEA indicates improved electron-accepting capability.

To further examine the chemical stability and reactivity of the clusters, we analyze the HOMO-LUMO energy gap and related global reactivity descriptors.
The trends observed in the VIP and VEA are further supported by reactivity and stability descriptors, including the highest occupied molecular orbital (HOMO) and lowest unoccupied molecular orbital (LUMO) gap, chemical hardness ($\eta$), chemical potential ($\mu$), electrophilicity index and nucleophilicity index, as summarized in table~\ref{tab:cdft}.
Chemical reactivity can be interpreted in terms of the response of the total electronic energy to changes in the number of electrons while keeping the external potential constant.
The first derivative of the total energy with respect to the number of electrons defines the chemical potential $\mu$:
\begin{equation}
\mu = \left( \frac{\partial E}{\partial N} \right)_v
\end{equation}
where, $E$ represents the total electronic energy, $N$ is the number of electrons, and $v$ denotes the external potential generated by the nuclei. 
Physically, the chemical potential describes the tendency of electrons to escape from a system and is closely related to electronegativity.
Within the frontier molecular orbital approximation, the chemical potential can be estimated using the energies of the HOMO and the LUMO orbitals. 
The difference between these two energy levels is defined as the HOMO–LUMO gap.
A larger gap means the cluster is more stable and less reactive, while a smaller gap indicates higher reactivity.

\begin{equation}
E_g = E_{\text{LUMO}} - E_{\text{HOMO}}
\end{equation}
and the chemical potential may be approximated as
\begin{equation}
\mu \approx \frac{1}{2}\left(E_{\text{HOMO}} + E_{\text{LUMO}}\right).
\end{equation}

Another important global descriptor is the chemical hardness $\eta$, which measures the resistance of a system to charge transfer or deformation of the electron cloud. 
It is defined as the second derivative of the total energy with respect to the number of electrons:
\begin{equation}
\eta = \left( \frac{\partial^2 E}{\partial N^2} \right)_v .
\end{equation}
Using the Koopman-type approximation, the chemical hardness can be estimated directly from the frontier orbital energies, as
\begin{equation}
\eta \approx \frac{1}{2}\left (E_{\text{LUMO}} - E_{\text{HOMO}}\right) .
\end{equation}
The systems with larger hardness values are generally more stable and less chemically reactive. 
Using the chemical potential and hardness, the electrophilicity index $\omega$ can be defined as
\begin{equation}
\omega = \frac{\mu^2}{2\eta}.
\end{equation}

This quantity represents the stabilization energy gained by a system when it acquires additional electronic charge and therefore provides a useful measure of the electron-accepting capability of the cluster.

\begin{table}[H]
\footnotesize
\caption{Reactivity and stability descriptors of the \wxby clusters, including the highest occupied molecular orbital (HOMO) and lowest unoccupied molecular orbital (LUMO) energy gap, chemical hardness $(\eta)$, chemical potential $(\mu)$, electrophilicity index $(\omega)$, and nucleophilicity index.}
\centering

\setlength{\tabcolsep}{4pt}
\renewcommand{\arraystretch}{1.1}

\begin{tabular}{
l 
c 
c 
c 
>{\centering\arraybackslash}m{2.0cm}
>{\centering\arraybackslash}m{2.0cm}
>{\centering\arraybackslash}m{2.1cm}
>{\centering\arraybackslash}m{2.1cm}
}
\toprule

\textbf{Structure} & 
\textbf{HOMO} & 
\textbf{LUMO} & 
\makecell{\textbf{Energy Gap}\\(eV)} & 
\makecell{\textbf{Chemical} \\ \textbf{Hardness}\\($\eta$, eV)} & 
\makecell{\textbf{Chemical}\\ \textbf{Potential}\\($\mu$, eV)} & 
\shortstack{\textbf{Electrophilicity}\\\textbf{Index}} &
\shortstack{\textbf{Nucleophilicity}\\\textbf{Index}} \\
\midrule

\textbf{\wthree}       & -5.21 & -2.84 & 2.37 & 1.18 & -4.03 & 1.54 & 4.72 \\
\textbf{\wtwobone}     & -5.65 & -2.90 & 2.76 & 1.38 & -4.28 & 1.39 & 4.15 \\
\textbf{\wonebtwo}     & -6.42 & -3.17 & 3.25 & 1.63 & -4.79 & 1.37 & 3.16 \\

\midrule

\textbf{\wfour}        & -4.63 & -2.56 & 2.07 & 1.02 & -3.59 & 1.35 & 4.50 \\
\textbf{\wthreebone}   & -5.15 & -2.67 & 2.48 & 1.24 & -3.91 & 1.38 & 4.18 \\
\textbf{\wtwobtwo}     & -5.34 & -2.99 & 2.35 & 1.18 & -4.16 & 1.53 & 3.79 \\
\textbf{\wonebthree}   & -6.37 & -2.89 & 3.48 & 1.74 & -4.63 & 1.50 & 3.54 \\

\midrule

\textbf{\wfive}        & -4.27 & -2.50 & 1.76 & 0.88 & -3.39 & 1.40 & 4.86 \\
\textbf{\wfourbone}    & -4.38 & -2.64 & 1.74 & 0.87 & -3.51 & 1.39 & 5.04 \\
\textbf{\wthreebtwo}   & -4.86 & -2.63 & 2.23 & 1.11 & -3.75 & 1.38 & 4.26 \\
\textbf{\wtwobthree}   & -5.45 & -3.20 & 2.25 & 1.13 & -4.33 & 1.73 & 3.82 \\
\textbf{\wonebfour}    & -6.66 & -3.09 & 3.57 & 1.79 & -4.88 & 1.83 & 3.45 \\

\bottomrule
\end{tabular}

\label{tab:cdft}
\end{table}

For pure tungsten clusters, the HOMO–LUMO gap is relatively small.
This is because tungsten has many $d$-electrons, which create closely spaced energy levels. 
As a result, these clusters show more metallic behavior and higher reactivity.
When boron is added, the HOMO–LUMO gap increases stabilizing the clusters and making them less reactive. 
For example, in the three-atom series, the energy gap increases from 2.37~eV (W$_3$) to 2.76~eV (W$_2$B$_1$) and further to 3.25~eV (W$_1$B$_2$). 
A similar trend is seen in larger clusters, where the gap increases from 2.07~eV (W$_4$) to 3.48~eV (W$_1$B$_3$), and from 1.76~eV (W$_5$) to 3.57~eV (W$_1$B$_4$). 
With boron addition, the separation between the occupied and unoccupied electronic states increases, which causes the gap to widen.
The chemical hardness ($\eta$) is also found to increases with boron concentration. 
For example, $\eta$ increases from 1.18~eV for W$_3$ to 1.63~eV for W$_1$B$_2$, while in the five-atom series it rises from 0.88~eV for W$_5$ to 1.79~eV for W$_1$B$_4$. 
The chemical potential ($\mu$) also follows these observations by presenting a negative value and monotonically decreasing with the increase of boron concentration.
For example, $\mu$ shifts from $-4.03$~eV (W$_3$) to $-4.79$~eV (W$_1$B$_2$), and from $-3.39$~eV (W$_5$) to $-4.88$~eV (W$_1$B$_4$). 
This shift indicates stronger electron binding and reduced tendency for electron donation as boron concentration increases.
The more negative chemical potential also supports a reduced chemical reactivity inferred by other descriptors.

The electrophilicity index~($\omega$) shows a moderate increase with boron substitution, particularly in boron-rich clusters.
For example, $\omega$ increases from 1.40~(W$_5$) to 1.83~(W$_1$B$_4$), indicating an enhanced tendency to accept electrons.
In particular, clusters such as W$_2$B$_3$ ($\omega = 1.73$) and W$_1$B$_4$ exhibit greater electrophilicity, suggesting a strong tendency to accept electrons.
On the contrary the nucleophilicity index decreases by increasing boron content.
For instance, it decreases from 4.72~(W$_3$) to 3.16~(W$_1$B$_2$) and from 4.86~(W$_5$) to 3.45~(W$_1$B$_4$). 
This reduction indicates a lower tendency to donate electrons, consistent with the stronger electron binding observed in boron-rich clusters.
The increase in HOMO-LUMO gap, chemical hardness, and VIP indicates enhanced electronic stability and stronger electron binding with boron incorporation. 
At the same time, the increase in electrophilicity and a reduction in nucleophilicity suggest an increasing tendency toward electron acceptance in boron-rich clusters.

\subsection{Vibrational properties}
\begin{figure}[H]
    \centering
    \includegraphics[width=\columnwidth]{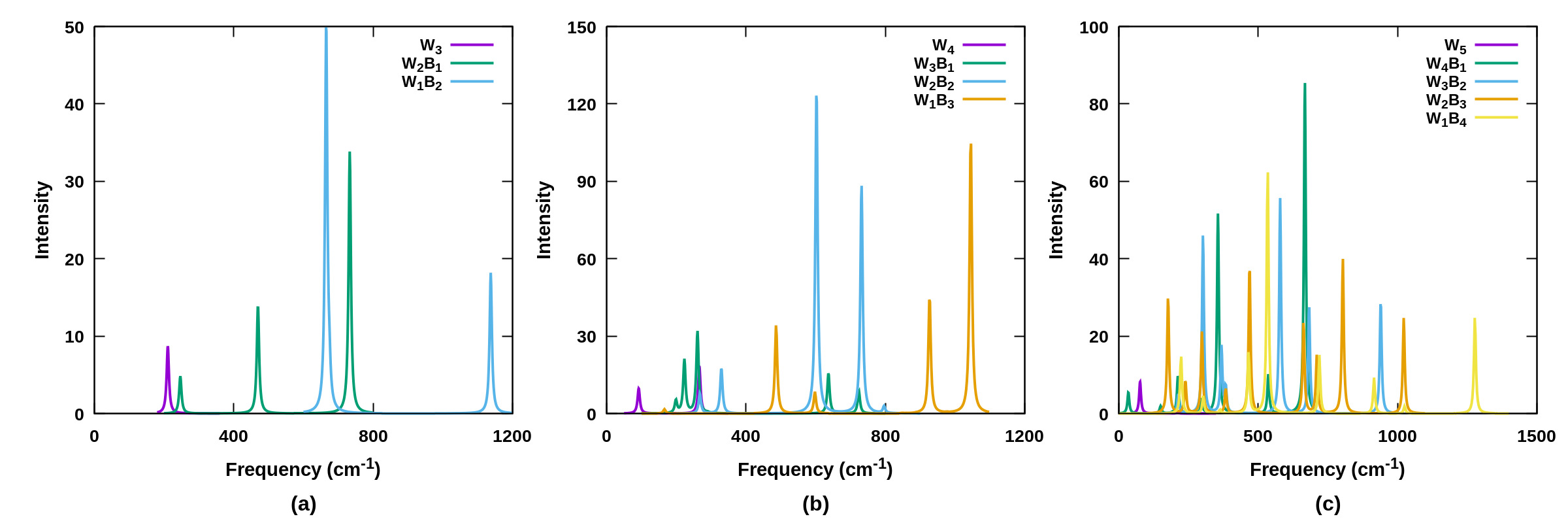}
    \caption{Vibrational spectra of a)~\wthree, b)~\wfour, and c)~\wfive~clusters with boron substitution.}
    \label{fig:ir_absorbance}
\end{figure}

In addition to structural and electronic properties, the vibrational behavior of the clusters provides important information about the nature and strength of bonding in the system.
The vibrational properties of W$_n$B$_m$ clusters were analyzed using calculated IR and Raman spectra. 
We first discuss the W$_3$ cluster and its boron-substituted derivatives (W$_2$B$_1$ and W$_1$B$_2$), followed by the corresponding trends in four and five atoms based clusters. 
All optimized structures exhibit only real vibrational frequencies, confirming their dynamical stability. 
The calculated IR and Raman active modes are summarized in SI~Table~1, while the overall spectral evolution with boron substitution is shown in Fig.~\ref{fig:ir_absorbance}.

Pure tungsten clusters~(W$_3$-W$_5$) mainly exhibit low-frequency vibrational modes below $\sim$370~cm$^{-1}$, originating from W-W stretching and bending vibrations.
These low-frequency modes arise from the large atomic mass of tungsten and the relatively soft metallic bonding within the cluster framework.
Upon boron substitution, additional modes appear in the intermediate frequency region ($\sim$350-800~cm$^{-1}$), which are primarily associated with the W-B stretching modes.
For example, W$_2$B$_1$ shows strong W-B stretching modes near 469 and 733~cm$^{-1}$ (SI~Table~1), indicating a stronger and more directional bonding compared to that in pure W-W clusters.

With increasing boron concentration, high-frequency modes associated with the B-B stretching modes become increasingly prominent.
For instance, W$_1$B$_2$ exhibits an intense B-B stretching mode near 1138~cm$^{-1}$, reflecting the formation of strong covalent B-B bonds.
In mixed clusters such as W$_3$B$_1$ and W$_3$B$_2$, the W-W, W-B, and B-B vibrational modes coexist, producing broader spectral distributions across low, intermediate, and high-frequency regions.
This behavior is consistent with our observation of coexistence of metallic and covalent bonding features in these intermediate compositions.

In boron-rich clusters such as W$_1$B$_3$ and W$_1$B$_4$, the spectra are dominated by multiple B-B stretching modes extending upto $\sim$1278~cm$^{-1}$.
The increasing intensity and multiplicity of these high-frequency modes indicate enhanced covalent character and stronger B-B interactions within the cluster, while the contribution from W-W vibrations becomes comparatively weaker.
The Raman spectra follow similar trends to that of the IR spectra, with strong Raman-active modes observed for W-B and B-B stretching vibrations in boron-rich clusters.
In particular, the intense Raman peaks near 796~cm$^{-1}$ for W$_2$B$_2$ and 1278~cm$^{-1}$ for W$_1$B$_4$ support the increasing covalent character introduced by boron incorporation (see SI~Table~1).

Overall, the vibrational spectra reveal a gradual shift from low-frequency modes in pure tungsten clusters to higher-frequency vibrational modes in boron-rich systems.
The emergence of distinct W-B and B-B stretching modes with increasing boron concentration reflects significant changes in the vibrational characteristics of the clusters.

\section{Summary and Conclusion}
\label{sec:concl}
In this work, the structural, electronic, and vibrational properties of small tungsten–boron clusters were systematically investigated using density functional theory (DFT) calculations employing the B3LYP exchange–correlation functional with the QZVP basis set. To explore possible structural configurations, the USPEX code interfaced with Gaussian was used to generate several isomeric structures for each cluster composition. From these generated configurations, the lowest-energy isomers were selected for detailed property calculations. The main objective of the study was to understand how the addition of boron atoms modifies the bonding, stability, and electronic behavior of tungsten clusters at the nanoscale.

The calculated structures show that pure tungsten clusters mainly prefer compact metallic geometries. However, when boron atoms are introduced, noticeable structural changes occur because boron atoms are smaller and form stronger directional bonds. Boron atoms preferentially occupy edge or outer positions in the clusters and gradually form B-B linkages as their concentration increases. Consequently, the clusters become less symmetric and more distorted, with shorter W-B and B-B bond lengths compared to W-W bonds. 

These structural modifications strongly influence the electronic properties and stability of the clusters. The calculated binding energies show that boron addition improves the overall stability of the clusters. At the same time, the HOMO–LUMO energy gap and chemical hardness increase with increasing boron concentration, indicating that boron-rich clusters become electronically more stable with the reduction of chemical reactivity. Similarly, the increase in ionization potential together with the more negative chemical potential values suggests stronger electron binding in boron-containing clusters. 

The changes in bonding and electronic structure are also reflected in the eigenvalue spectra and vibrational behavior of the clusters. Boron substitution increases the separation between occupied and unoccupied electronic states, leading to enhanced electronic stabilization. In addition, the vibrational spectra gradually shift from low-frequency W-W vibrational modes in pure tungsten clusters to higher-frequency W-B and B-B vibrational modes in boron-rich systems. The appearance of these high-frequency modes reflects the increasing covalent character introduced by boron atoms. 

Overall, the results show that the addition of boron gradually increases the covalent character of tungsten clusters through the formation of stronger and more directional W–B and B b bonds. The improved stability of boron-rich clusters is mainly due to the formation of stronger W-B and B-B bonds, which also enhance chemical hardness and reduce chemical reactivity. Thus, boron acts as an important stabilizing element by significantly modifying the structural and electronic properties of tungsten clusters. These findings provide atomic-level insight into the bonding behavior and stability of tungsten–boron systems and may help in understanding the fundamental properties of boron-rich tungsten-based nanostructures and related tungsten boride materials.

\section{Acknowledgments}
One of us (B.J.N.) gratefully acknowledges the National PARAM Supercomputing Facility (NPSF), C-DAC, Pune, India, and PARAM Rudra at the Inter-University Accelerator Centre (IUAC), New Delhi, for providing high-performance computing resources and technical support for this work. The author also acknowledges the financial support from the Anusandhan National Research Foundation (ANRF) under the sanctioned project “ANRF PAIR IITB Hub” (Sanction Order No. ANRF/PAIR/2025/000017/PAIR).

\section*{Credit authorship contribution statement}
\textbf{Akshata M. Waghmare:} Data curation, Formal analysis, Investigation, Software, Validation, Visualization, Writing-original draft, Writing-review and editing.
\textbf{Gautam S. Kamat:} Formal analysis, Investigation, Validation, Writing-review and editing.
\textbf{Sajeev S. Chacko:} Data curation, Formal analysis, Investigation, Methodology, Validation, Visualization, Writing-original draft, Writing-review and editing.
\textbf{Chetan V. Gurada:} Project Supervision, Writing-original draft, Writing-review and editing.
\textbf{Balasaheb J. Nagare:} Conceptualization, Data curation, Formal analysis, Investigation, Methodology, Project administration, Supervision, Validation, Visualization, Writing-original draft, Writing-review and editing.

\section*{Declaration of competing interest}
The authors declare that they have no known competing financial interests or personal relationships that could have appeared to influence the work reported in this paper.

\section*{Data Availability}
The data that support the findings of this study are available from the corresponding author upon reasonable request.

\printbibliography

@article{kim2019synthesis,
  title={Synthesis of metal boride nanoparticles using triple thermal plasma jet system},
  author={Kim, Minseok and Oh, Jeong-Hwan and Kim, Tae-Hee and Lee, Yong Hee and Hong, Seung-Hyun and Choi, Sooseok},
  journal={Journal of Nanoscience and Nanotechnology},
  volume={19},
  number={10},
  pages={6264--6270},
  year={2019},
  publisher={American Scientific Publishers},
  doi={10.1166/jnn.2019.17026}
}

@article{moscicki2021properties,
  title={Properties of spark plasma sintered compacts and magnetron sputtered coatings made from Cr, Mo, Re and Zr alloyed tungsten diboride},
  author={Mo{\'s}cicki, Tomasz and Psiuk, Rafa{\l} and Radziejewska, Joanna and Wi{\'s}niewska, Maria and Garbiec, Dariusz},
  journal={Coatings},
  volume={11},
  number={11},
  pages={1378},
  year={2021},
  publisher={MDPI}
}

@article{fuger2022anisotropic,
  title={Anisotropic super-hardness of hexagonal WB$_{2\pm z}$ thin films},
  author={Fuger, Christoph and Hahn, Rainer and Zauner, Lukas and Wojcik, Tomasz and Weiss, M. and Limbeck, A. and Hunold, O. and Polcik, P. and Riedl, H.},
  journal={Materials Research Letters},
  volume={10},
  number={2},
  pages={70--77},
  year={2022},
  publisher={Taylor \& Francis},
  doi={10.1080/21663831.2021.2021308}
}

@article{psiuk2023mechanical,
  title={Mechanical and thermal properties of W-Ta-B coatings deposited by high-power impulse magnetron sputtering (HiPIMS)},
  author={Psiuk, Rafa{\l} and Mo{\'s}cicki, Tomasz and Chrzanowska-Gi{\.z}y{\'n}ska, Justyna and Kurpaska, {\L}ukasz and Radziejewska, Joanna and Denis, Piotr and Garbiec, Dariusz and Chmielewski, Marcin},
  journal={Materials},
  volume={16},
  number={2},
  pages={664},
  year={2023},
  publisher={MDPI},
  doi={10.3390/ma16020664}
}

@article{radziejewska2020characterization,
  title={Characterization and wear response of magnetron sputtered W--B and W--Ti--B coatings on WC--Co tools},
  author={Radziejewska, Joanna and Psiuk, Rafa{\l} and Mo{\'s}cicki, Tomasz},
  journal={Coatings},
  volume={10},
  number={12},
  pages={1231},
  year={2020},
  publisher={MDPI},
  doi={10.3390/coatings10121231}
}

@article{fuger2019influence,
title = {Influence of Tantalum on Phase Stability and Mechanical Properties of WB$_2$},
  author={Fuger, Christoph and Moraes, Vincent and Hahn, Rainer and Bolvardi, Hamid and Polcik, Peter and Riedl, Helmut and Mayrhofer, Paul Heinz},
  journal={MRS Communications},
  volume={9},
  number={1},
  pages={375--380},
  year={2019},
  publisher={Cambridge University Press},
  doi={10.1557/mrc.2019.5}
}

@article{park2019high,
  title={High-current-density HER electrocatalysts: graphene-like boron layer and tungsten as key ingredients in metal diborides},
  author={Park, Hyounmyung and Zhang, Yuemei and Lee, Eunsoo and Shankhari, Pritam and Fokwa, Boniface PT},
  journal={ChemSusChem},
  volume={12},
  number={16},
  pages={3726--3731},
  year={2019},
  publisher={Wiley Online Library},
  doi={10.1002/cssc.201901301}
}

@article{levine2009advancements,
  title={Advancements in the search for superhard ultra-incompressible metal borides},
  author={Levine, Jonathan B. and Tolbert, Sarah H. and Kaner, Richard B.},
  journal={Advanced Functional Materials},
  volume={19},
  number={22},
  pages={3519--3533},
  year={2009},
  publisher={Wiley Online Library},
  doi={10.1002/adfm.200901257}
}

@article{wu2026facilitated,
  title={Facilitated Polysulfide Redox Conversion by Delocalized Electrons in MBene Heterointerface for Highly Stable Lithium--Sulfur Batteries},
  author={Wu, Guifen and Fan, Yunmiao and Li, Jiatong and Shen, Zhaoxi and Xie, Yuxiu and Yang, Peixun and Pu, Jun},
  journal={Nano-Micro Letters},
  volume={18},
  number={1},
  pages={252},
  year={2026},
  publisher={Springer},
  doi={10.1007/s40820-026-02100-3}
}

@article{moscicki2020influence,
  title={Influence of overstoichiometric boron and titanium addition on the properties of RF magnetron sputtered tungsten borides},
  author={Moscicki, Tomasz and Psiuk, Rafal and S{\l}omi{\'n}ska, Hanna and Levintant-Zayonts, Neonila and Garbiec, Dariusz and Pisarek, Marcin and Bazarnik, Piotr and Nosewicz, Szymon and Chrzanowska-Gi{\.z}y{\'n}ska, Justyna},
  journal={Surface and Coatings Technology},
  volume={390},
  pages={125689},
  year={2020},
  publisher={Elsevier},
  doi={10.1016/j.surfcoat.2020.125689}
}

@article{gu2022synthesis,
  title={Synthesis, phase evolutions, and stabilities of boron-rich tungsten borides at high pressure},
  author={Gu, Chao and Zhou, Xuefeng and Ma, Dejiang and Zhao, Yusheng and Wang, Shanmin},
  journal={Inorganic Chemistry},
  volume={61},
  number={45},
  pages={18193--18200},
  year={2022},
  publisher={ACS Publications},
  doi={10.1021/acs.inorgchem.2c02957}
}

@article{weigend2005balanced,
  title={Balanced basis sets of split valence, triple zeta valence and quadruple zeta valence quality for H to Rn: Design and assessment of accuracy},
  author={Weigend, Florian and Ahlrichs, Reinhart},
  journal={Physical Chemistry Chemical Physics},
  volume={7},
  number={18},
  pages={3297--3305},
  year={2005},
  publisher={Royal Society of Chemistry},
  doi={10.1039/B508541A}
}

@article{lee1988development,
  title={Development of the Colle-Salvetti correlation-energy formula into a functional of the electron density},
  author={Lee, Chengteh and Yang, Weitao and Parr, Robert G},
  journal={Physical review B},
  volume={37},
  number={2},
  pages={785},
  year={1988},
  publisher={APS},
  doi={10.1103/PhysRevB.37.785}
}

@article{becke1993density,
  title={Density-functional thermochemistry. III. The role of exact exchange},
  author={Becke, Axel D},
  journal={The Journal of chemical physics},
  volume={98},
  number={7},
  pages={5648--5652},
  year={1993},
  publisher={American Institute of Physics},
  doi={10.1063/1.464913}
}

@article{hu1992optical,
  title={Optical and Raman spectroscopy of mass-selected tungsten dimers in argon matrices},
  author={Hu, Zhendong and Dong, Jian-Guo and Lombardi, John R. and Lindsay, D. M.},
  journal={The Journal of chemical physics},
  volume={97},
  number={11},
  pages={8811--8812},
  year={1992},
  publisher={American Institute of Physics},
  doi={10.1063/1.463353}
}

@article{sharma2022dft,
  title = {{A {DFT} study of {Se$_n$Te$_n$} clusters}},
  author={Sharma, Tamanna and Sharma, Raman and Kanhere, D. G.},
  journal={Nanoscale Advances},
  volume={4},
  number={5},
  pages={1464--1482},
  year={2022},
  publisher={Royal Society of Chemistry},
  doi={10.1039/D1NA00321F}
}

@article{kiessling1947crystal,
  title={The crystal structures of molybdenum and tungsten borides},
  author={Kiessling, Roland and Wetterholm, Allan and Sill{\'e}n, Lars Gunnar and Linnasalmi, Annikki and Laukkanen, Pentti},
  journal={Acta Chem. Scand},
  volume={1},
  number={10},
  pages={893--916},
  year={1947},
  doi={10.3891/ACTA.CHEM.SCAND.01-0893}
}

@article{armas1973chemical,
  title={Chemical vapour deposition of molybdenum and tungsten borides by thermal decomposition of gaseous mixtures of halides on a solar “front chaud”},
  author={Armas, B. and Trombe, F.},
  journal={Solar Energy},
  volume={15},
  number={1},
  pages={67--73},
  year={1973},
  publisher={Elsevier},
  doi={10.1016/0038-092X(73)90008-X}
}

@article{itoh1987formation,
  title={Formation process of tungsten borides by solid state reaction between tungsten and amorphous boron},
  author={Itoh, H. and Matsudaira, T. and Naka, S. and Hamamoto, H. and Obayashi, M.},
  journal={Journal of materials science},
  volume={22},
  pages={2811--2815},
  year={1987},
  publisher={Springer},
  doi={10.1007/BF01086475}
}

@article{qin2018effect,
  title = {{Effect of Pressure on the Structural, Electronic and Mechanical Properties of Ultraincompressible {W$_2$B}}},
  author={Qin, Zhen and Gong, Weiguang and Song, Xianqi and Wang, Menglong and Wang, Hongbo and Li, Quan},
  journal={RSC advances},
  volume={8},
  number={62},
  pages={35664--35671},
  year={2018},
  publisher={Royal Society of Chemistry},
  doi={10.1039/C8RA05706K}
}

@article{zhong2013phase,
  title = {{Phase Stability, Physical Properties, and Hardness of Transition-Metal Diborides {MB$_2$} ({M = Tc, W, Re, Os}): First-Principles Investigations}},
  author={Zhong, Ming-Min and Kuang, Xiao-Yu and Wang, Zhen-Hua and Shao, Peng and Ding, Li-Ping and Huang, Xiao-Fen},
  journal={The Journal of Physical Chemistry C},
  volume={117},
  number={20},
  pages={10643--10652},
  year={2013},
  publisher={ACS Publications},
  doi={10.1021/jp400204c}
}

@article{zhao2010phase,
  title={Phase stability and mechanical properties of tungsten borides from first principles calculations},
  author={Zhao, Erjun and Meng, Jian and Ma, Yanming and Wu, Zhijian},
  journal={Physical Chemistry Chemical Physics},
  volume={12},
  number={40},
  pages={13158--13165},
  year={2010},
  publisher={Royal Society of Chemistry},
  doi={10.1039/C004122J}
}

@article{chen2008electronic,
  title = {Electronic and Structural Origin of Ultraincompressibility of {5$d$} Transition-Metal Diborides {MB$_2$} ({M = W, Re, Os})},
  author={Chen, Xing-Qiu and Fu, Chong Long and Kr{\v{c}}mar, M. and Painter, Gayle S.},
  journal={Physical review letters},
  volume={100},
  number={19},
  pages={196403},
  year={2008},
  publisher={APS},
  doi={10.1103/PhysRevLett.100.196403}
}

@article{zhang2017first,
  title={First-Principles predictions of phase transition and mechanical properties of tungsten diboride under pressure},
  author={Zhang, Huai-Yong and Xi, Feng and Zeng, Zhao-Yi and Chen, Xiang-Rong and Cai, Ling-Cang},
  journal={The Journal of Physical Chemistry C},
  volume={121},
  number={13},
  pages={7397--7403},
  year={2017},
  publisher={ACS Publications},
  doi={10.1021/acs.jpcc.6b12621}
}

@article{zhao2018unexpected,
  title={Unexpected stable phases of tungsten borides},
  author={Zhao, Changming and Duan, Yifeng and Gao, Jie and Liu, Wenjie and Dong, Haiming and Dong, Huafeng and Zhang, Dekun and Oganov, Artem R},
  journal={Physical Chemistry Chemical Physics},
  volume={20},
  number={38},
  pages={24665--24670},
  year={2018},
  publisher={Royal Society of Chemistry},
  doi={10.1039/C8CP04222E}
}

@article{kvashnin2020wb5,
  title = {{W{B}$_{5-x}$}: Synthesis, Properties, and Crystal Structure-New Insights into the Long-Debated Compound},
  author={Kvashnin, Alexander G. and Rybkovskiy, Dmitry V. and Filonenko, Vladimir P. and Bugakov, Vasilii I. and Zibrov, Igor P. and Brazhkin, Vadim V. and Oganov, Artem R. and Osiptsov, Andrey A. and Zakirov, Artem Ya.},
  journal={Advanced Science},
  volume={7},
  number={16},
  pages={2000775},
  year={2020},
  publisher={Wiley Online Library},
  doi={10.1002/advs.202000775}
}

@article{windsor2021tungsten,
  title={Tungsten boride shields in a spherical tokamak fusion power plant},
  author={Windsor, Colin G. and Astbury, Jack O. and Davidson, James J. and McFadzean, Charles J.R. and Morgan, J. Guy and Wilson, Christopher L. and Humphry-Baker, Samuel A.},
  journal={Nuclear Fusion},
  volume={61},
  number={8},
  pages={086018},
  year={2021},
  publisher={IOP Publishing},
  doi={10.1088/1741-4326/ac09ce}
}

@article{duschanek1995critical,
  title = {{Critical Assessment and Thermodynamic Calculation of the Binary System {(B-W)}}},
  author={Duschanek, H. and Rogl, Po.},
  journal={Journal of phase equilibria},
  volume={16},
  pages={150--161},
  year={1995},
  publisher={Springer},
  doi={10.1007/BF02664852}
}

@article{zhang2005density,
  title = {{Density Functional Theory Study of {W$_n$} ($n = 2$-$4$) Clusters}},
  author={Zhang, Xiurong and Ding, Xunlei and Dai, Bing and Yang, Jinlong},
  journal={Journal of Molecular Structure: THEOCHEM},
  volume={757},
  number={1-3},
  pages={113--118},
  year={2005},
  publisher={Elsevier},
  doi={10.1016/j.theochem.2005.09.021}
}

@misc{g03,
author = {M. J. Frisch and G. W. Trucks and H. B. Schlegel and G. E. Scuseria and
M. A. Robb and J. R. Cheeseman and Montgomery, Jr., J. A. and T. Vreven and
K. N. Kudin and J. C. Burant and J. M. Millam and S. S. Iyengar and J. Tomasi and
V. Barone and B. Mennucci and M. Cossi and G. Scalmani and N. Rega and
G. A. Petersson and H. Nakatsuji and M. Hada and M. Ehara and K. Toyota and
R. Fukuda and J. Hasegawa and M. Ishida and T. Nakajima and Y. Honda and O. Kitao and
H. Nakai and M. Klene and X. Li and J. E. Knox and H. P. Hratchian and J. B. Cross and
V. Bakken and C. Adamo and J. Jaramillo and R. Gomperts and R. E. Stratmann and
O. Yazyev and A. J. Austin and R. Cammi and C. Pomelli and J. W. Ochterski and
P. Y. Ayala and K. Morokuma and G. A. Voth and P. Salvador and J. J. Dannenberg and
V. G. Zakrzewski and S. Dapprich and A. D. Daniels and M. C. Strain and
O. Farkas and D. K. Malick and A. D. Rabuck and K. Raghavachari and
J. B. Foresman and J. V. Ortiz and Q. Cui and A. G. Baboul and S. Clifford and
J. Cioslowski and B. B. Stefanov and G. Liu and A. Liashenko and P. Piskorz and
I. Komaromi and R. L. Martin and D. J. Fox and T. Keith and M. A. Al-Laham and
C. Y. Peng and A. Nanayakkara and M. Challacombe and P. M. W. Gill and
B. Johnson and W. Chen and M. W. Wong and C. Gonzalez and J. A. Pople},
title = {Gaussian 03, \uppercase{R}evision \uppercase{C}.02},
note = {\uppercase{G}aussian, Inc., Wallingford, CT, 2004},
}

@article{cumberland2005osmium,
  title={Osmium diboride, an ultra-incompressible, hard material},
  author={Cumberland, Robert W. and Weinberger, Michelle B. and Gilman, John J. and Clark, Simon M. and Tolbert, Sarah H. and Kaner, Richard B.},
  journal={Journal of the American Chemical Society},
  volume={127},
  number={20},
  pages={7264--7265},
  year={2005},
  publisher={ACS Publications},
  doi={10.1021/ja043806y}
}

@article{yin2013theory,
  title = {Theoretical Study on Polarizability and Dipole Moment of {W$_m$B$_n$} ({$m+n\le7$}) Clusters},
  author={Yin, Lin and Zhang, Xiu Rong and Zhang, Fu Xing},
  journal={Advanced Materials Research},
  volume={774},
  pages={479--483},
  year={2013},
  publisher={Trans Tech Publ},
  doi={10.4028/www.scientific.net/AMR.774-776.479}
}

@article{Xiuron2013DensityFT,
  title = {Density Functional Theory Study of Structure and Stability of {$W_mB_n$} Clusters ({$m+n \leq 7$})},
  author = {Zhang, Xiurong and Yin, Lin and Li, Weijun and Wang, Yangyang and Yuan, Aihua},
  journal = {Chinese Journal of Computational Physics},
  volume = {30},
  number = {5},
  pages = {775--782},
  year = {2013},
  url={https://api.semanticscholar.org/CorpusID:101116294}
}

@article{lin2014structural,
  title={Structural and electronic properties of tungsten nanoclusters by DFT and basin-hopping calculations},
  author={Lin, Ken-Huang and Wang, Shi-Liang and Chen, Chuan and Ju, Shin-Pon},
  journal={RSC advances},
  volume={4},
  number={46},
  pages={24286--24294},
  year={2014},
  publisher={Royal Society of Chemistry},
  doi={10.1039/C4RA02053G}
}

@article{yong2009first,
  title={First principles study of neutral and charged small tungsten clusters},
  author={Yong, Xu and Xian-Long, Wang and Zhi, Zeng},
  journal={Acta Physica Sinica},
  volume={58},
  number={6},
  pages={S72--S78},
  year={2009},
  publisher={CHINESE PHYSICAL SOC PO BOX 603, BEIJING 100080, PEOPLES R CHINA},
  doi={10.7498/aps.58.72}
}

@article{lin2008first,
  title = {First-Principles Calculations on the Structures of {$W_n$} ({$n = 3$--$27$}) Clusters},
  author = {Lin, Qiu-Bao and Li, Ren-Quan and Wen, Yu-Hua and Zhu, Zi-Zhong},
  journal = {Acta Physica Sinica},
  volume = {57},
  number = {1},
  pages = {181--185},
  year = {2008},
  doi = {10.7498/aps.57.181}
}

@article{du2009geometrical,
 title = {{Geometrical and electronic structures of small {W$_n$} ({$n = 2$--$16$}) clusters}},
  author={Du, Jiguang and Sun, Xiyuan and Meng, Daqiao and Zhang, Pengcheng and Jiang, Gang},
  journal={The Journal of chemical physics},
  volume={131},
  number={4},
  year={2009},
  publisher={AIP Publishing},
  doi={10.1063/1.3187525}
}

@article{wang2005geometries,
  title = {{Geometries and electronic structures of {W$_4$} and {W$_n$} clusters}},
  author={Wang, J. P. and Niu, C. J. and Wu, Z. J.},
  journal={International journal of quantum chemistry},
  volume={101},
  number={3},
  pages={334--339},
  year={2005},
  publisher={Wiley Online Library}
}

@article{qin2008rhenium,
  title={Is rhenium diboride a superhard material?},
  author={Qin, Jiaqian and He, Duanwei and Wang, Jianghua and Fang, Leiming and Lei, Li and Li, Yongjun and Hu, Juan and Kou, Zili and Bi, Yan},
  journal={Advanced Materials},
  volume={20},
  number={24},
  pages={4780--4783},
  year={2008},
  publisher={WILEY-VCH Verlag Weinheim},
  doi={10.1002/adma.200801471}
}

@article{erdogan2025enhanced,
  title={Enhanced radiation shielding performance of tungsten borides-epoxy composites},
  author={Erdogan, Furkan and Bermudez, Santiago and Mohammadi, Reza and Rojas, Jessika V.},
  journal={Composites Science and Technology},
  pages={111233},
  year={2025},
  publisher={Elsevier},
  doi={10.1016/j.compscitech.2025.111233}
}

@article{liu2023boriding,
  title={Boriding of tungsten by the powder-pack process: Phase formation, growth kinetics and enhanced neutron shielding},
  author={Liu, Yue and Liu, Xiaoyi and Lai, Chen and Ma, Jie and Meng, Xianfang and Zhang, Long and Xu, Guanglong and Lu, Yiwen and Li, Hongyi and Wang, Jinshu and  Chen, Shuqun},
  journal={International Journal of Refractory Metals and Hard Materials},
  volume={110},
  pages={106049},
  year={2023},
  publisher={Elsevier},
  doi={10.1016/j.ijrmhm.2022.106049}
}

@article{windsor2022activation,
  title={Activation and transmutation of tungsten boride shields in a spherical tokamak},
  author={Windsor, Colin G. and Astbury, Jack O. and Morgan, J. Guy and Wilson, Christopher L. and Humphry-Baker, Samuel A.},
  journal={Nuclear Fusion},
  volume={62},
  number={3},
  pages={036009},
  year={2022},
  publisher={IOP Publishing},
  doi={10.1088/1741-4326/ac4866}
}

@article{lin2022oxidation,
  title = {Oxidation Resistance of {WB} and {W$_2$B--W} Neutron Shields},
  author={Lin, Yusha and McFadzean, Charles and Humphry-Baker, Samuel A},
  journal={Journal of Nuclear Materials},
  volume={565},
  pages={153762},
  year={2022},
  publisher={Elsevier},
  doi={10.1016/j.jnucmat.2022.153762}
}

@article{liu2025understanding,
  title={Understanding the friction and wear behavior of tungsten boride coating at temperatures of 400-800° C},
  author={Liu, Yue and Xu, Lu and Shao, Mengxue and Qi, Jiahui and Rainforth, W. Mark and Chen, Shuqun},
  journal={Wear},
  pages={206413},
  year={2025},
  publisher={Elsevier},
  doi={10.1016/j.wear.2025.206413}
}

@article{mishigdorzhiyn2020microstructure,
  title={Microstructure and wear behavior of tungsten hot-work steel after boriding and boroaluminizing},
  author={Mishigdorzhiyn, Undrakh and Chen, Yan and Ulakhanov, Nikolay and Liang, Hong},
  journal={Lubricants},
  volume={8},
  number={3},
  pages={26},
  year={2020},
  publisher={MDPI},
  doi={10.3390/lubricants8030026}
}

@article{artamonov1966abrasive,
  title={Abrasive power of refractory compounds},
  author={Artamonov, A. Ya. and Bezykornov, A. I. and Ivanov, A. N.},
  journal={Soviet Powder Metallurgy and Metal Ceramics},
  volume={5},
  number={9},
  pages={722--725},
  year={1966},
  publisher={Springer},
  doi={10.1007/BF00774099}
}

@article{zhang2023tungsten,
  title = {Tungsten Boride Stabilized Single-Crystal {LiNi$_{0.83}$Co$_{0.07}$Mn$_{0.1}$O$_2$} Cathode for High Energy Density Lithium-Ion Batteries: Performance and Mechanisms},
  author={Zhang, Qimeng and Deng, Qiang and Zhong, Wentao and Li, Jing and Wang, Ziming and Dong, Pengyuan and Huang, Kevin and Yang, Chenghao},
  journal={Advanced Functional Materials},
  volume={33},
  number={27},
  pages={2301336},
  year={2023},
  publisher={Wiley Online Library},
  doi={10.1002/adfm.202301336}
}

@article{hatipoglu2022design,
  title={Design of metal-substituted tungsten diboride as an efficient bifunctional electrocatalyst for hydrogen and oxygen evolution},
  author={Hatipoglu, Ezgi and Peighambardoust, Naeimeh Sadat and Sadeghi, Ebrahim and Aydemir, Umut},
  journal={International Journal of Energy Research},
  volume={46},
  number={12},
  pages={17540--17555},
  year={2022},
  publisher={Wiley Online Library},
  doi={10.1002/er.8421}
}

@misc{nist_b2,
  title={NIST Computational Chemistry Comparison and Benchmark Database: B$_2$ Experimental Data},
  author={{NIST}},
  url={https://cccbdb.nist.gov}
}

@article{borin2010electronic,
  title={Electronic structure and chemical bonding in W$_2$ molecule},
  author={Borin, Antonio Carlos and Gobbo, Jo{\~a}o Paulo and Roos, Bj{\"o}rn O},
  journal={Chemical Physics Letters},
  volume={490},
  number={1-3},
  pages={24--28},
  year={2010},
  publisher={Elsevier},
  doi={10.1016/j.cplett.2010.03.022}
}

@article{merriles2021chemical,
  title={Chemical bonding and electronic structure of the early transition metal borides: ScB, TiB, VB, YB, ZrB, NbB, LaB, HfB, TaB, and WB},
  author={Merriles, Dakota M. and Nielson, Christopher and Tieu, Erick and Morse, Michael D.},
  journal={The Journal of Physical Chemistry A},
  volume={125},
  number={20},
  pages={4420--4434},
  year={2021},
  publisher={ACS Publications},
  doi={10.1021/acs.jpca.1c02886}
}

@article{yanagisawa2000investigation,
  title={An investigation of density functionals: The first-row transition metal dimer calculations},
  author={Yanagisawa, Susumu and Tsuneda, Takao and Hirao, Kimihiko},
  journal={The Journal of Chemical Physics},
  volume={112},
  number={2},
  pages={545--553},
  year={2000},
  publisher={American Institute of Physics},
  doi={10.1063/1.480546}
}

@article{barden2000homonuclear,
  title={Homonuclear 3 d transition-metal diatomics: A systematic density functional theory study},
  author={Barden, Christopher J. and Rienstra-Kiracofe, Jonathan C. and Schaefer III, Henry F.},
  journal={The Journal of Chemical Physics},
  volume={113},
  number={2},
  pages={690--700},
  year={2000},
  publisher={American Institute of Physics},
  doi={10.1063/1.481916}
}

@article{chertovskikh2021study,
  title={Study of {TiB}$_2$ coated hard alloy tool wear resistance during titanium alloy machining},
  author={Chertovskikh, S. V. and Shuster, L. Sh. and Fox-Rabinovich, G. S.},
  journal={Chemical and Petroleum Engineering},
  volume={57},
  number={7},
  pages={690--695},
  year={2021},
  publisher={Springer},
  doi={10.1007/s10556-021-00993-y}
}
\end{document}